\documentclass[twocolumn,showpacs]{revtex4}
\usepackage{graphicx}
\usepackage{epstopdf}
\usepackage{bm}

\begin{document}
\title{Dynamical control of matter-wave tunneling in periodic potentials}

\author{H. Lignier, C. Sias, D. Ciampini, Y. Singh, A. Zenesini, O. Morsch and E. Arimondo}

\affiliation{CNR-INFM, Dipartimento di Fisica `E. Fermi',
Universit\`{a} di Pisa, Largo Pontecorvo 3, 56127 Pisa, Italy}

\begin{abstract}
We report on measurements of dynamical suppression of inter-well
tunneling of a Bose-Einstein condensate (BEC) in a strongly driven
optical lattice. The strong driving is a sinusoidal shaking of the
lattice corresponding to a time-varying linear potential, and the
tunneling is measured by letting the BEC freely expand in the
lattice. The measured tunneling rate is reduced and, for certain
values of the shaking parameter, completely suppressed. Our
results are in excellent agreement with theoretical predictions.
Furthermore, we have verified that in general the strong shaking
does not destroy the phase coherence of the BEC, opening up the
possibility of realizing quantum phase transitions by using the
shaking strength as the control parameter.
\end{abstract}

 \maketitle

Quantum tunneling of particles between potential wells connected
by a barrier is a fundamental physical effect. While typically
quantum systems decay faster when they are perturbed, if the wells
are periodically shaken back and forth (or a time-varying
potential is applied in a different way), the tunneling rate can
actually be reduced and, for certain shaking strengths, even
completely suppressed~\cite{grossmann_91,eckardt_05}.

Modifications of the dynamics of quantum systems by applying
periodic potentials have been investigated in a number of contexts
including the renormalization of Land\'{e} $g$-factors in atoms
~\cite{haroche_70}, the micromotion of a single trapped
ion~\cite{raab00} and the motion of electrons in semiconductor
superlattices~\cite{keay_95}. In particular, theoretical studies
of double-well systems and of periodic potentials have led to the
closely related concepts of coherent destruction of tunneling and
dynamical localization~\cite{dunlap_86,grossmann_91}. In the
latter, tunneling between the sites of a periodic array is
inhibited by applying a periodically varying potential, e.g. by
shaking the array back and forth (see Fig. 1), and as a
consequence the tunneling parameter $J$ representing the gain in
kinetic energy in a tunneling event is replaced by
$|J_{\mathrm{eff}}|<|J|$. In a number of experiments signatures of
this tunneling suppression have been
observed~\cite{keay_95,madison_98,iyer_07}, and recently dynamical
localization and coherent suppression of tunneling have been
demonstrated using light propagating in coupled waveguide
arrays~\cite{longhi_06,dellavalle_07}. Also, the predictions of
the Bose-Hubbard model in a moving frame were recently
tested~\cite{mun07}. So far, however, an exact experimental
realization of the intrinsically nonlinear Bose-Hubbard
model~\cite{eckardt_05} driven by a time-periodic potential has
not been reported.

\begin{figure}[ht]
\includegraphics[width=9cm]{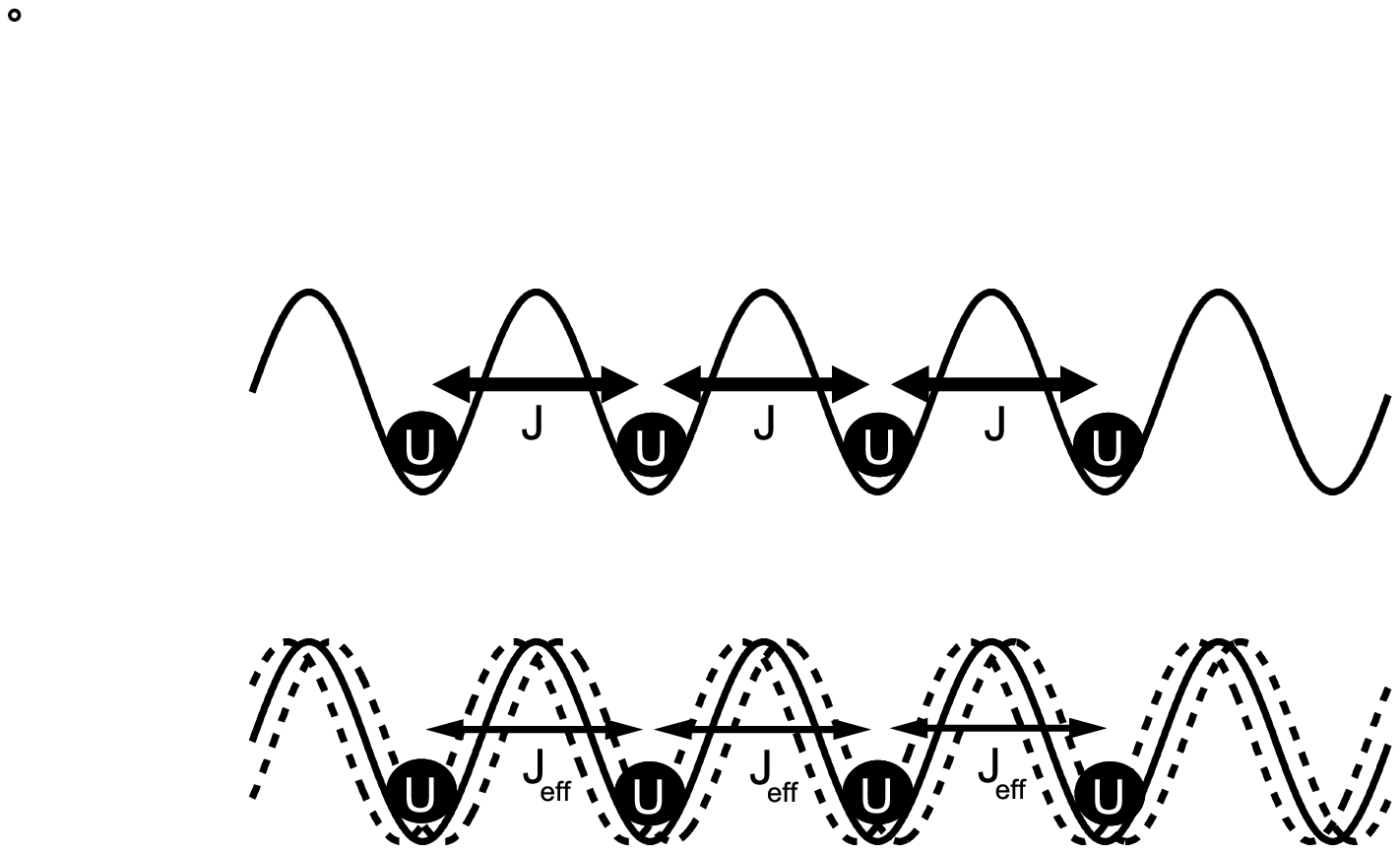}
\caption{\label{figure1} Suppression of tunneling by strong
driving. The dynamics of a Bose-Einstein condensate in a periodic
potential is governed by the tunneling matrix element $J$ and the
on-site interaction energy $U$ (above). If the potential is
strongly shaken, tunneling between the wells is dynamically
suppressed, leading to a renormalized tunneling matrix element
$J_\mathrm{eff}$ (below) but leaving the interaction energy $U$
unaffected.}
\end{figure}

In this Letter, we report on the observation of the dynamical
tunneling suppression predicted in
refs.~\cite{eckardt_05,creffield_06} using Bose-Einstein
condensates (BECs) in strongly driven periodic optical
potentials~\cite{morsch_review}. In contrast to other systems, the
characteristics of such optical lattices - potential depth,
lattice spacing, driving strength and frequency - can be freely
chosen and allow us to control the tunneling over a wide range of
parameters. In this way we were able to experimentally confirm
theoretical predictions with great accuracy. Also, our system
allows us to observe the effects of the shaking both by monitoring
the real-space expansion of the BEC in the optical lattice and by
performing time-of-flight experiments in which the phase coherence
of the BEC can be measured. The latter experiments allow us to
verify that the tunneling suppression occurs in a phase-coherent
way in spite of the strong shaking.

Furthermore, BECs have an intrinsic nonlinear on-site interaction
energy (represented by $U$ in Fig. 1), the interplay of which with
the tunneling parameter $J$ has been shown to lead to the
Mott-insulator quantum phase transition for a critical value of
the ratio $U/J$~\cite{jaksch_98,greiner_02a}. It has been
theoretically predicted that for a BEC in a shaken optical
lattice, this ratio can be replaced by $U/J_{\mathrm{eff}}$ and
hence that it should be possible to drive the system across the
quantum phase transition by varying the shaking
parameter~\cite{eckardt_05,creffield_06}. In this work, we
demonstrate the feasibility of the key ingredients of this scheme.
In particular, we show that when tunneling in the shaken lattice
is completely suppressed, the phase coherence of the BEC is lost
in agreement with the physical picture of a sudden `switch-off' of
the inter-well coupling and a subsequent independent evolution of
the local phases due to collisions between the atoms
~\cite{greiner_02b,Li_07}.

Our system consisting of a Bose-Einstein condensate inside a
(sinusoidally) shaken one-dimensional optical lattice is
approximately described by the Hamiltonian
\begin{equation}
\hat{H}_0=-J\sum_{\langle
i,j\rangle}(\hat{c}_i^\dagger\hat{c}_j+\hat{c}_j^\dagger\hat{c}_i)+\frac{U}{2}\sum_j
\hat{n}_j(\hat{n}_j-1)+K\cos(\omega t)\sum_j j\hat{n}_j,
\end{equation}
where $\hat{c}_i^{(\dagger)}$ are the boson creation and
annihilation operators on site $i$,
$\hat{n}_i=\hat{c}_i^{\dagger}\hat{c}_i$ are the number operators,
and $K$ and $\omega$ are the strength and angular frequency of the
shaking, respectively. The first two terms in the Hamiltonian
describe the Bose-Hubbard model~\cite{jaksch_98} with the
tunneling matrix element $J$ and the on-site interaction term $U$.
The shaking of the lattice is expected to renormalize the
tunneling matrix element $J$, leading to an effective tunneling
parameter~\cite{eckardt_05}
\begin{equation}
J_\mathrm{eff}=J\mathcal{J}_0(K_0),
\label{bessel}
\end{equation}
where $\mathcal{J}_0$ is the zeroth-order ordinary Bessel function
and we have introduced the dimensionless parameter
$K_0=K/\hbar\omega$.

In our experiment we created BECs of about $5\times 10^4$
87-rubidium atoms using a hybrid approach in which evaporative
cooling was initially effected in a magnetic time-orbiting
potential (TOP) trap and subsequently in a crossed dipole trap.
The dipole trap was realized using two intersecting gaussian laser
beams at $1030\,\mathrm{nm}$ wavelength and a power of around
$1\,\mathrm{W}$ per beam focused to waists of $50\,\mathrm{\mu
m}$. After obtaining pure condensates of around $5\times 10^4$
atoms the powers of the trap beams were adjusted in order to
obtain elongated condensates with the desired trap frequencies
($\approx 20\,\mathrm{Hz}$ in the longitudinal direction and
$80\,\mathrm{Hz}$ radially). Along the axis of one of the dipole
trap beams a one-dimensional optical lattice potential was then
added by ramping up the power of the lattice beams in
$50\,\mathrm{ms}$ (the ramping time being chosen such as to avoid
excitations of the BEC). The optical lattices used in our
experiments were created using two counter-propagating gaussian
laser beams ($\lambda = 852\,\mathrm{nm}$) with $120\,\mathrm{\mu
m}$ waist and a resulting optical lattice spacing $d_L= \lambda /2
= 0.426 \,\mathrm{\mu m}$. The depth $V_0$ of the resulting
periodic potential is measured in units of $E_{\rm rec}= \hbar^2
\pi^2 / (2m d_L^2)$, where $m$ is the mass of the Rb atoms. By
introducing a frequency difference $\Delta \nu$ between the two
lattice beams (using acousto-optic modulators which also control
the power of the beams), the optical lattice could be moved at a
velocity $v=d_L\Delta \nu$ or accelerated with an acceleration
$a=d_L\frac{d\Delta\nu}{dt}$. In order to periodically shake the
lattice, $\Delta \nu$ was sinusoidally modulated with angular
frequency $\omega$, leading to a time-varying velocity $v(t)=
d_L\Delta\nu_{\mathrm{max}}\sin(\omega t)$ and hence to a
time-varying force
\begin{equation}
F(t)= m\omega d_L \Delta\nu_{\mathrm{max}}\cos(\omega t)=
F_{\mathrm{max}}\cos(\omega t).
\end{equation}
The peak shaking force $F_{\mathrm{max}}$ is related to the
shaking strength $K$ appearing in Eq. (1) by
\begin{equation}
K=F_{\mathrm{max}}d_L,
\end{equation}
 and hence the dimensionless shaking
parameter
\begin{equation}
K_0=\frac{K}{\hbar \omega} =
\frac{md_L^2\Delta\nu_{\mathrm{max}}}{\hbar}=\frac{\pi^2\Delta\nu_{\mathrm{max}}}{2\omega_{\mathrm{rec}}}.
\end{equation}
The spatial shaking amplitude $\Delta x_{\mathrm{max}}$ can then
be written as
\begin{equation}
\Delta
x_{\mathrm{max}}=\frac{2}{\pi^2}\frac{\omega_{\mathrm{rec}}}{\omega}K_0 d_L,
\end{equation}
so for a typical shaking frequency $\omega/2\pi = 3\,\mathrm{kHz}$
we have $\Delta x_{\mathrm{max}}\approx 0.5 d_L$ at $K_0=2.4$.

After loading the BECs into the optical lattice, the frequency
modulation of one of the lattice beams creating the shaking was
switched on either suddenly or using a linear ramp with a
timescale of a few milliseconds. Finally, in order to measure the
effective tunneling rate $|J_{\mathrm{eff}}|$ between the lattice
wells (where the modulus indicates that we are not sensitive to
the sign of $J$, in contrast to the time-of-flight experiments
described below), we then switched off the dipole trap beam that
confined the BEC along the direction of the optical lattice,
leaving only the radially confining beam switched on (the trap
frequency of that beam along the lattice direction was on the
order of a few Hz and hence negligible on the timescales of our
expansion experiments, which were typically less than
$200\,\mathrm{ms}$). The BEC was now free to expand along the
lattice direction through inter-well tunneling and its {\it
in-situ} width was measured using a resonant flash, the shadow
cast by which was imaged onto a CCD chip. The observed density
distribution was then fitted with one or two gaussians.

In a preliminary experiment without shaking ($K_0=0$), we verified
that for our expansion times the growth in the condensate width
$\sigma_x$ along the lattice direction was to a good approximation
linear and that the dependence of $d\sigma_x/dt$ on the lattice
depth (up to $V_0/E_{\mathrm{rec}}=9$) followed the expression for
$J(V_0/E_\mathrm{rec})$ in the lowest energy
band~\cite{zwerger_03}
\begin{equation}
J\left(\frac{V_0}{E_\mathrm{rec}}\right)= \frac{4 E_\mathrm{rec}
}{\sqrt{\pi}}\left( \frac{V_0}{E_{\mathrm{rec}}} \right)^{3/4}
e^{-2\sqrt{V_0/E_\mathrm{rec}}.}\end{equation} This enabled us to
confirm that $d\sigma_x/dt$ measured at a fixed time was directly
related to $J$ and, in a shaken lattice, to
$|J_{\mathrm{eff}}(K_0)|$. The results of our measurements of
$|J_{\mathrm{eff}}(K_0)/J|$,
 for various
lattice depths $V_0$ and driving frequencies $\omega$ are
summarized in Fig. 2. We found a universal behaviour of
$|J_{\mathrm{eff}}/J|$ that is in very good agreement with the
Bessel-function re-scaling of Eq. (\ref{bessel}). We were able to
measure $|J_{\mathrm{eff}}/J|$ for $K_0$ up to $12$, albeit
agreement with theory beyond $K_0 \approx 6$ was  less good, with
the experimental values lying consistently below the theoretical
curve. For the zeroes of the $\mathcal{J}_0$ Bessel function at
$K_0 \approx 2.4$ and $5.4$, complete suppression of tunneling was
observed (within our experimental resolution, we could measure a
suppression by at least a factor of $25$).

\begin{figure}[ht]
\includegraphics[width=9cm]{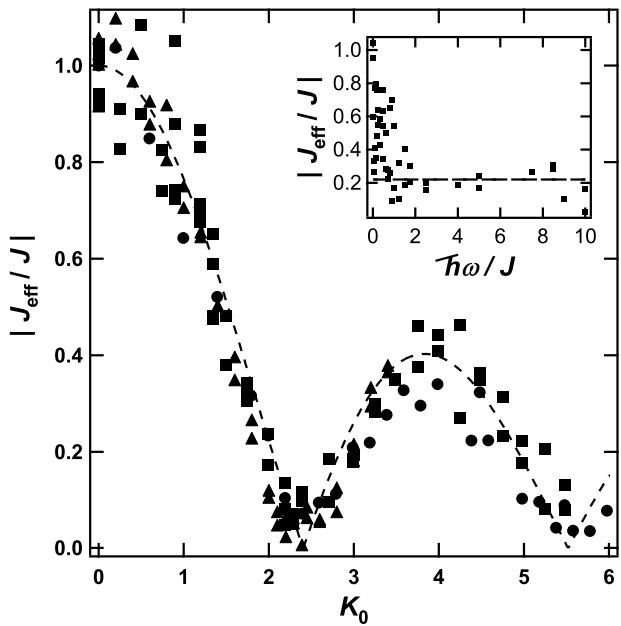}
\caption{\label{figure2} Dynamical suppression of tunneling in an
optical lattice. Shown here are the values of
$|J_{\mathrm{eff}}/J|$ as calculated from the expansion velocities
as a function of the shaking parameter $K_0$. The lattice depths
and shaking frequencies were: $V_0/E_{\mathrm{rec}}=6$,
$\omega/2\pi = 1\,\mathrm{kHz}$ (squares),
$V_0/E_{\mathrm{rec}}=6$, $\omega/2\pi = 0.5\,\mathrm{kHz}$
(circles), and $V_0/E_{\mathrm{rec}}=4$, $\omega/2\pi =
1\,\mathrm{kHz}$ (triangles). The dashed line is the theoretical
prediction. {\it Insert:} Dependence of the tunneling suppression
$|J_{\mathrm{eff}}/J|$ on the shaking frequency $\omega$ for
$K_0=2.0$ and $V_0/E_\mathrm{rec}=9$ corresponding to $J/h=90$
Hz.}
\end{figure}

We also checked the behaviour of $|J_{\mathrm{eff}}/J|$ as a
function of $\omega$ for a fixed value of $K_0=2$ (see insert in
Fig. 2) and found that over a wide range of frequencies between
$\hbar \omega/J\approx 0.3$ and $\hbar \omega/J\approx 30$ the
tunneling suppression due to the shaking of the lattice works,
although for $\hbar \omega/J\lesssim 1$ we found that
$|J_{\mathrm{eff}}(K_0)/J|$ as a function of $K_0$ deviated from
the Bessel function near the zero points, where the suppression
was less efficient than expected. In the limit of large shaking
frequencies ($\omega/2\pi \gtrsim 3\,\mathrm{kHz}$, to be compared
with the typical mean separation of $ \approx 15\,\mathrm{kHz}$
between the two lowest two energy bands at $V_0/E_{\rm rec} =9$),
we observed excitations of the condensate to the first excited
band of the lattice. In our {\it in-situ} expansion measurements,
these band excitations were visible in the condensate profile as a
broad gaussian pedestal below the near-gaussian profile of the
ground-state condensate atoms. From the widths of those pedestals
we inferred that $|J_{\mathrm{eff}}/J|$ of the atoms in the
excited band also followed the Bessel-function rescaling of Eq.
(\ref{bessel}), and that the ratios of the tunneling rates in the
two bands agreed with theoretical models.

\begin{figure}[ht]
\includegraphics[width=9cm]{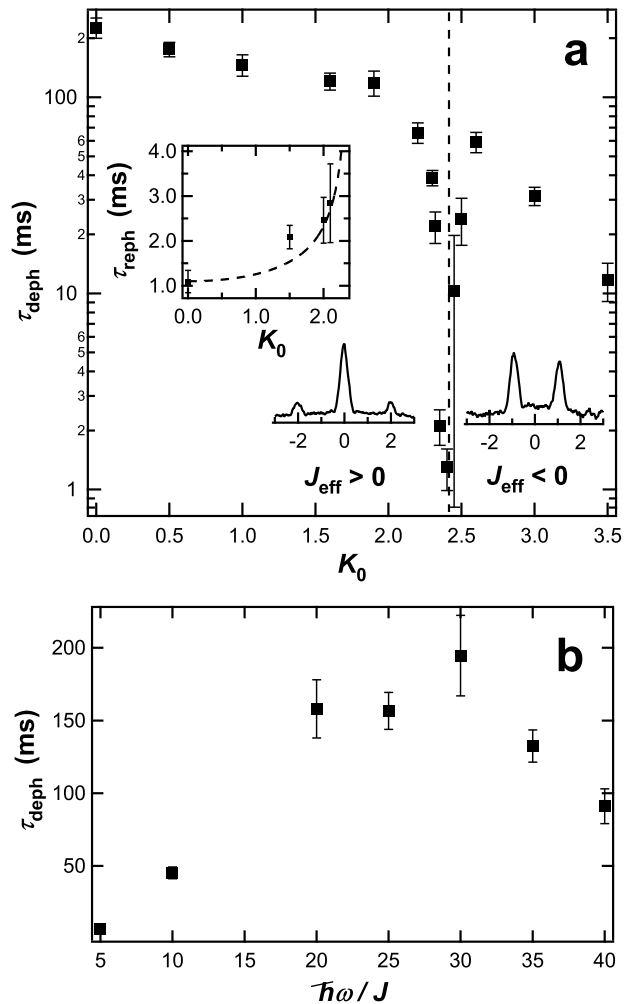}
\caption{\label{figure3}Phase coherence in a shaken lattice. (a)
Dephasing time $\tau_{\mathrm{deph}}$ (decay time of the
visibility) of the condensate as a function of $K_0$ for a lattice
with $V_0/E_\mathrm{rec}=9$ and $\omega/2\pi = 3\,\mathrm{kHz}$.
The vertical dashed line marks the position of $K_0=2.4$ dividing
the regions with $J_\mathrm{eff}>0$ (left) and $J_\mathrm{eff}<0$
(right). In both regions, a typical (vertically integrated)
interference pattern of a time-of-flight experiment without final
acceleration to the zone edge is shown (on the $x$-axis, the
spatial position has been converted into the corresponding
momentum in units of the recoil momentum $p_\mathrm{rec}=h/d_L$.)
{\it Insert:} Rephasing time after dephasing at $K_0=2.4$ and
subsequent reduction of $K_0$. (b) Dephasing time as a function of
the normalized driving frequency $\hbar\omega/J$ for $K_0=2.2$.}
\end{figure}

We now turn to the phase coherence of the BEC in the shaken
lattice. In order to quantify the degree of phase coherence, after
shaking the condensate in the lattice for a fixed time between $1$
and $\approx 200\,\mathrm{ms}$ we accelerated the lattice for
$\approx 1\,\mathrm{ms}$ so that at the end of the acceleration
the BEC was in a staggered state at the edge of the Brillouin
zone. After switching off the dipole trap and lattice beams and
letting the BEC fall under gravity for $20\,\mathrm{ms}$, this
resulted in an interference pattern featuring two peaks of roughly
equal height~\cite{morsch_decay}. In the region between the first
two zeroes of the Bessel function, where $\mathcal{J}_0<0$, we
found an interference pattern (see Fig. 3 (a)) that was shifted by
half a Brillouin zone, in agreement with theoretical predictions.
We then measured the visibility
$\mathcal{V}=(h_{\mathrm{max}}-h_{\mathrm{min}})/(h_{\mathrm{max}}+h_{\mathrm{min}})$
of the interference pattern as a function of the time the
condensate spent inside the shaken lattice, where
$h_{\mathrm{max}}$ is the mean value of the condensate density at
the position of the two interference peaks and $h_{\mathrm{min}}$
is the condensate density in a region of width equal to about
$1/4$ of the peak separation centered about the halfway point
between the two peaks. For a perfectly phase-coherent condensate
$\mathcal{V}\approx 1$, whereas for a strongly dephased condensate
$\mathcal{V}\approx 0$. For $K_0 \lesssim 2.2$, the BEC phase
coherence was maintained for several tens of milliseconds,
demonstrating that the tunneling could be suppressed by a factor
of up to $10$ over hundreds of shaking cycles without
significantly disturbing the BEC. This result is expressed more
quantitatively in Fig. 3 (a). Here, the condensate was held in the
lattice ($V_0/E_{\mathrm{rec}}=9$), and the shaking was switched
on suddenly at $t=0$ (we found no significantly different
behaviour when $K_0$ was linearly ramped in a few milliseconds).
Thereafter, the visibility was measured as a function of time and
the decay time constant $\tau_{\rm deph}$ of the resulting
near-exponential function was extracted. Apart from a slow overall
decrease in the dephasing time for increasing $K_0$, a sharp dip
around $K_0=2.4$ is visible. In this region, $J$ is suppressed by
a factor of more than $20$ and hence the effective tunneling rate
$|J_\mathrm{eff}/h| \lesssim 10\,\mathrm{Hz}$, which for our
experimental parameters is comparable to the on-site interaction
energy $U$ expressed in frequency units (we checked that the
widths of the on-site wavefunctions and hence $U$ were independent
of $K_0$ by analyzing the side-peaks in the interference pattern).
This means that neighbouring lattice sites are effectively
decoupled and the local phases evolve independently due to
interatomic collisions, leading to a dephasing of the
array~\cite{greiner_02a,Li_07}. By increasing the dipole trap
frequency (and hence $U$), we verified that the timescale for this
dephasing decreases as expected. We also studied a re-phasing of
the BEC when, after an initial dephasing at $K_0=2.4$, the value
of the shaking parameter was reduced below $2.4$. The time
constant $\tau_{\mathrm{reph}}$ of the subsequent rephasing of the
condensate (mediated by inter-well tunneling and on-site
collisions) increased with decreasing $J_{\mathrm{eff}}$ (see the
insert of Fig. 3 (a), where we compare $\tau_{\mathrm{reph}}$ as a
function of $K_0$ with the inverse of the generalized Josephson
frequency $\omega_{\mathrm{Josephson}}^{-1}\propto
J_{\mathrm{eff}}^{-1/2}$ predicted by the two-well
model~\cite{smerzi_97,Li_07}).

Finally, we investigated the dependence of the dephasing time on
the shaking frequency $\omega$ (see Fig. 3 (b)). Interestingly,
while the tunneling suppression as observed {\it } in-situ works
even for $\hbar\omega/J\approx 1$, in order to maintain the phase
coherence of the condensate, much larger shaking frequencies are
needed. Indeed, for our system there exists an optimum shaking
frequency of $\hbar\omega/J\approx 30$.

In summary, we have measured the dynamical suppression of
tunneling of a BEC in strongly shaken optical lattices and found
excellent agreement with theoretical predictions. Our results show
that the tunneling suppression occurs in a phase-coherent way and
can, therefore, be used as a tool to control the tunnelling matrix
element while leaving the on-site interaction energy unchanged (in
contrast to the usual technique of increasing the lattice depth,
which changes both) and without disturbing the condensate. This
might ultimately lead to the possibility of controlling quantum
phase transitions by strong driving of the lattice. In this
context, it will be important to investigate the question of
adiabaticity when dynamically changing the shaking parameter.
Furthermore, our system also opens up other avenues of research
such as the realization of exact dynamical localization using
discontinuous shaking waveforms~\cite{dignam_02,iyer_07} or
tunneling suppression in superlattices~\cite{creffield_07}.

This work was supported by OLAQUI and MIUR-PRIN. The authors would
like to thank Sandro Wimberger for useful discussions.

\bibliographystyle{apsrmp}

\end{document}